\documentclass[twocolumn,showpacs,preprintnumbers,amsmath,amssymby,superscriptaddress]{revtex4}

\usepackage{dcolumn}
\usepackage{bm}
\usepackage{slashed}
\usepackage[export]{adjustbox}
\usepackage[utf8]{inputenc}
\usepackage{graphicx,xcolor}
\usepackage{amsmath, amsfonts, amssymb, bm}
\usepackage{float}
\usepackage{tikz}

\begin{document}

\preprint{APS/123-QED}

\title{ Four-photon Kapitza-Dirac Effect as Electron Spin Filter \\}

\author{Asma Ebadati}
\affiliation{Laser-Plasma Research Institute, Shahid Beheshti University, G. C., Evin, Tehran, I. R. Iran}
\author{Mohsen Vafaee}
\affiliation{Department of Chemistry, Tarbiat Modares University, P. O. Box 14115-175, Tehran, I. R. Iran}
\author{Babak Shokri}
\email{ b.shokri@sbu.ac.ir}
\affiliation{Laser-Plasma Research Institute, Shahid Beheshti University, G. C., Evin, Tehran, I. R. Iran}

\date{\today}

\begin{abstract}
We theoretically demonstrate the feasibility of producing electron beam splitter using Kapitza-Dirac diffraction on bichromatic standing waves which are created by the fundamental frequency and the third harmonic. The relativistic electron in Bragg regime absorbs three photons with frequency of $\omega $ and emits a photon with frequency of $3\omega$, four-photon Kapitza-Dirac effect. In this four-photon Kapitza-Dirac effect distinct spin effects arise in different polarizations of the third harmonic laser beam. It is shown that the shape of Rabi oscillation between initial and scattered states is changed and finds two unequal peaks. In circular polarization for fundamental and third harmonic, despite Rabi oscillation, the spin down electron in 0.56 fs intervals maintains its momentum and spin. Also we present an electron spin filter with combination of a linearly polarized fundamental laser beam and a third harmonic with circular polarization that scatters the electron beam according to its spin state.  
\end{abstract}

\pacs{Valid PACS appear here}
\maketitle

\section{\label{sec:level1}Introduction\protect\\
}
The Stern-Gerlach experiment provided an evidence for existence of spin as an intrinsic, non-classical property [1]. A beam of silver atoms traveling through an inhomogeneous magnetic field is deflected up or down depending on their spin. Strangely, this experiment did not work with beams of electrons [2]. Bohr and Pauli emphasized, free electrons cannot be spin-polarized by exploiting magnetic fields, because of the combined effects of the Lorentz force and quantum uncertainty principle. This conclusion was due to the concept of classical particle trajectories and became a general argument in scientific literature [3-5].

One of the first efforts refuting the Bohr and Pauli’s statement was using a longitudinal magnetic-field configuration instead of the standard transverse geometry of Stern-Gerlach. In this way, the complete spin splitting with quantum-mechanical analysis has been reported [6-9]. In recent theoretical studies, the use of the grating and electromagnetic fields are resulted in the spin separation for electrons. Tang proposed a spin-polarized Talbot effect which is  non-paraxial for an electron beam scattered from a grating of magnetic nanostructures [10]. Also a transverse Stern-Gerlach magnet which diffracts electrons by a magnetic phase grating was discussed by McGregor \textit{et al} [11]. They indicated that by applying a current to the solenoids, a spin-dependent phase difference is created between the two arms of the Mach-Zehnder interferometer [12]. Moreover, a novel space-variant Wien filter, named "q-filter", which is composed of space variant orthogonal electric and magnetic fields can act as an efficient spin-polarization filter. This filter couples spin angular momentum to orbital angular momentum for electron, neutron, or atom beams [13-14]. 

Kapitza-Dirac (KD) effect, the quantum mechanical diffraction of an electron on a periodic spatial structure formed by a standing light wave, confirmed that it could be a way to detect the spin of electrons [15]. Ahrens \textit{et al.} theoretically showed complete spin-flip transitions applying relativistic treatment for KD effect [16-17]. It was demonstrated that electrons according to their spin state in the interaction with circularly polarized counter-propagating monochromatic standing waves can be separated [18]. In the meantime, Dellweg \textit{et al}. reported a similar way of generating spin-polarized electrons by using bichromatic ($\omega$:$2\omega$) laser beams in Kapitza-Dirac effect [19-20]. Also Dellweg showed that spin dynamics of electrons in bichromatic KD effect is dependent on the polarization of laser beams and therefore the spin direction of the output beams can be controlled [21-22]. 

One major result of these studies is that the initial electron with spin up state is transferred to the scattered electron with spin down and vice versa. This is symmetric flipping spin dynamics. For especial two cases, circularly polarization in KD effect with equal frequency [18] and combination of linear-circular polarization in bichromatic KD effect with a frequency ratio of 2  [22] the spin of electron does not flip and preserves its state. 

In the present paper, we theoretically discuss bichromatic KD effect arising from the interaction of the laser beam with frequency of $\omega$ and the counter-propagating laser beam with frequency of $3\omega$. In these fields, the electron exchanges four photons and the four-photon bichromatic KD (4PBKD) occurs. We investigate the spin polarization of the electron using various polarized counter propagating lasers beams with a frequency ratio of three. We focus on the electron whose momentum is parallel to the laser beam axis so that the interaction term $ \vec{p}.\vec{A} $ becomes very insignificant. Our article is organized as follows: in Sec. II we determine that in bichromatic standing waves with a frequency ratio of 1:3, electron exchanges four photons. Then based on the $ \textit{S} $ matrix approach, transition amplitude for resonant state is calculated and polarization dependent Rabi frequency is taken into account. Numerical solutions of the time-dependent Dirac equation in momentum space are applied to clear up relativistic quantum dynamics in Sec. III and look into scattering probability result of the different polarizations of two-color beams. In this paper atomic units are used throughout unless otherwise stated.

\section{Theoretical Description}
\subsection{\label{sec:level2}Electron Dispersion in Bragg Regime}
The interaction between an electron and two counter-propagating laser fields of frequency $ \omega_i(i=1,2)$ is due to the absorption of some photons from laser beam 1 and the  stimulated emission of some photons to laser beam 2. In the case of an intense bichromatic standing wave, the multiphoton interaction between electron and laser field becomes more likely. The absorption of $ N_a $ photons with $ \omega_1 $ frequency and emission of $ N_e $ photons with $ \omega_2 $ frequency conserve energy and momentum if $N_a\omega_1=N_e\omega_2$. In case of $ \omega_1=\omega $ and $ \omega_2=3\omega $, a free electron absorbs three  $ \omega $ photons and emits one $ 3\omega $ photon or vice versa. Since a photon has the energy $ c \hbar k  $ and the momentum $\hbar k  $, the total transferred energy is  $\triangle E=c (N_a k_1-N_e k_2) $ and the total transferred momentum is $ \triangle p= (N_ak_1+N_ek_2) $. In the presence of the fundamental frequency and the third harmonic with $ N_a=3$ and $N_e=1$, after the interaction the electron momentum change is $ 6k $.

The relativistic energy-momentum relation secant and quantum pathway that increases electron momentum by $ 6k $ are shown in Fig. 1. The total exchange of energy and momentum of the electron with laser beams is represented by horizontal dash line.
This slop is given by $ s=\frac{\triangle E}{\triangle p} $ and connects initial and final momenta of diffracted electron [17].
All pathways in Bragg regime starts and ends exactly on the dispersion relation secant. Theoretically, other transitions are also possible as well, but we focus on the resonant two-state quantum dynamics in Bragg regime. Absorption of one photon of $ 3\omega $ and emission of three photons of $ \omega $ make no difference in result.
\begin{figure}
\resizebox{86mm}{68mm}{\includegraphics{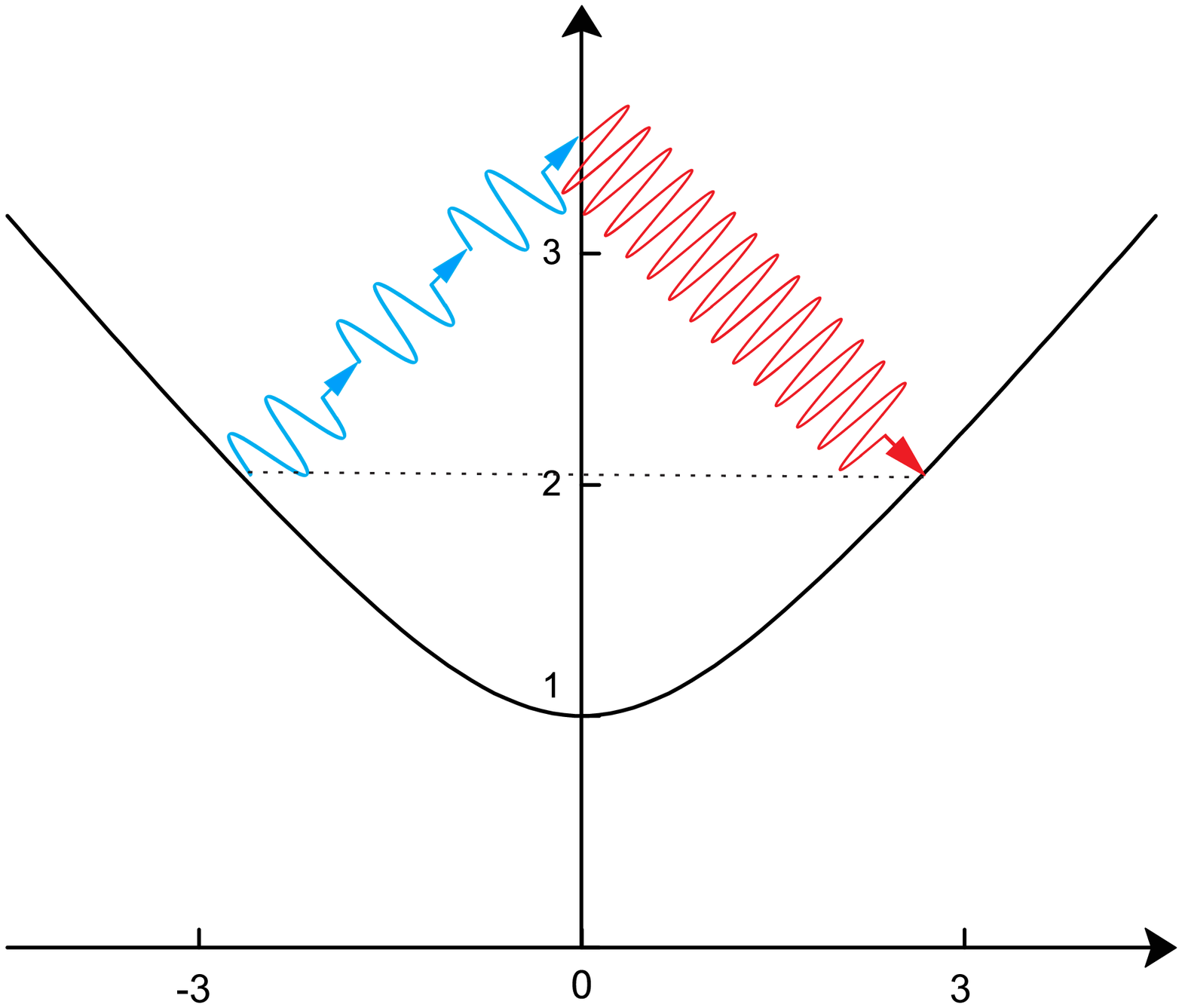}}
\begin{picture}(0.8,0.6)
\put(-14,25) {$\dfrac{p}{mc^2}$}
\end{picture}
\begin{picture}(0.7,0.1)
\put(5,200) {$\dfrac{\varepsilon(p)}{mc^2}$}
\end{picture}
\caption{\label{fig:epsart} Sketch of dominant pathway of four-photon KD effect in bichromatic standing waves. The dispersion relation of energy and momentum is in Bragg regime and each wiggly arrow shows a photon.}
\end{figure}
\subsection{\label{sec:level2}\textit{S} Matrix Approach}

The evaluation of the relativistic electron in four-photon Kapitza-Dirac process is described quantum mechanically by Dirac equation
\begin{equation}
\left( 
i\hbar\slashed{\partial}+\dfrac{e}{c}\slashed{A}(x)-m
 \right)\psi(x)=0,
\end{equation}
 in which $\slashed{A}  $ is denoted by Feynman slash $ \slashed{A}=\gamma.A $  and $\gamma$ stands for Dirac matrices.
 
  An analytical treatment for multiphoton stimulated Compton scattering such as bichromatic Kapitza-Dirac scattering has been accomplished through $ \textit{S} $ matrix approach with suitable approximations [22]. The Dirac equation has a well known solution; Volkov state for an electron in the case of the external potential being a plane wave. By using the fundamental laser mode and Volkov sates for the incident electron, the remaining four-photon Kapitza-Dirac process can be represented within the first order perturbation theory [23-24]. The calculation of S matrix is the same as that used in three photons KD effect, except that a further photon participates in the interaction.
  
In the presence of third harmonic the $ \textit{S} $ matrix for transition from $ p $ to $ p^\prime $ by absorbing three photons from $ A_1 $ as beam with fundamental frequency and emitting one photon into $ A_2 $ as beam with $ 3\omega$ frequency is given by 
  \begin{eqnarray}
    S &\approx& \frac{ie}{cV} \int d^4x\, \bar{u}_{p', s'} \Bigg(
     \slashed{A}_2^{(+)} \tilde{J}_2 e^{i \left( p' - p - 3k_1 \right) \cdot x} \nonumber\\
     & & \left. - \frac{e}{2 c} \left[ \frac{\slashed{A}_1^{(-)} \slashed{k} \slashed{A}_2^{(+)}}{k_1 \cdot p'} + \frac{\slashed{A}_2^{(+)} \slashed{k}_1 \slashed{A}_1^{(-)}}{k_1 \cdot p} \right]
     \tilde{J}_1 e^{i \left( p' - p - k_1 \right) \cdot x}
     \right) u_{p, s} \nonumber\\
     &\approx& \frac{i e}{2} T \bar{u}_{p', s'} \left[
      a_2 \tilde{J}_2 \bar{\slashed{\epsilon}}_2
     - \frac{e a_1 a_2}{4 c} \tilde{J}_1 \left( \frac{\slashed{\epsilon}_1 \slashed{k}_1 \bar{\slashed{\epsilon}}_2}{k_1 \cdot p'}
      + \frac{\bar{\slashed{\epsilon}}_2 \slashed{k}_1 \slashed{\epsilon}_1}{k_1 \cdot p} \right)
     \right] u_{p, s}. \nonumber\\
  \end{eqnarray}
  Here, $\slashed{A}_1^{(-)} = \frac{1}{2} a_1 \slashed{\epsilon}_1 e^{-i k_1 \cdot x}$ is the component that defines the absorption of one photon from $ A_1 $, and  $\slashed{A}_2^{(+)} = \frac{1}{2} a_2 \bar{\slashed{\epsilon}}_2 e^{i k_2 \cdot x}$  where $\bar{\slashed{\epsilon}}_2=\epsilon_2^*\cdot\gamma$ is the component that describes emission of one photon into $ A_2 $. Also $ \tilde{J}_{1,2}$ are generalized Bessel functions. In this derivation only resonant scattering process was considered, which fulfilled the Bragg condition. Therefore the $d^4x$- integration does result in the factor $c VT $, with $ T $ the interaction time and $ V$ quantification volume [22]. The initial and scattered electron momentum is set respectively to $p=(p^0,p_x,0,-3 \hbar k)$ and $p^\prime=(p^0,p_x,0,+3 \hbar k)$. The Dirac spinors for theses momenta are corresponding to Pauli spinors by
 \begin{equation}
u_{(p,s)}=\frac{1}{\sqrt{2mc(p^0+mc)}}\quad \begin{pmatrix} 
(p^0+mc)\chi_s \\
\vec{p}.\vec{\sigma}\chi_s
\end{pmatrix}.
\quad
\end{equation}
  We can calculate a part of Eq. (2) as
\begin{equation}
   \left( \bar{u}_{p', s'} \bar{\slashed{\epsilon}}_2 u_{p, s} \right)_{s', s}
    = -\frac{p_x}{m c} \vec{\epsilon}_2^{~*} \cdot \vec{e}_x + i \frac{3\omega}{m c^2} \left( \vec{\epsilon}_2^{~*}\times\vec{e}_z \right)\cdot \vec{\sigma}      
  \end{equation}
  and in a similar way
  \begin{eqnarray}
    & & \left( \bar{u}_{p',s'}\left[
     \frac{\slashed{\epsilon}_1 \slashed{k}_1 \bar{\slashed{\epsilon}}_2}{k_1 \cdot p'} + \frac{\bar{\slashed{\epsilon}}_2 \slashed{k}_1 \slashed{\epsilon}_1}{k_1 \cdot p}
    \right] u_{p,s} \right)_{s', s} \nonumber\\
 &\approx& \dfrac{2 \vec{\epsilon}_1 \cdot \vec{\epsilon_2^{~*}}}{m c} 
    + \frac{6 i \omega}{m^2 c^3} \left( \vec{\epsilon}_1 \times \vec{\epsilon_2}^{*} \right)\cdot \vec{\sigma}. 
  \end{eqnarray}
 Also from the Taylor series of the generalized Bessel functions, we can estimate 
   \begin{eqnarray}
   \tilde{J}_1 \approx -3\frac{e a_{1}}{m c^2} \frac{p_{x}}{m c} \vec{\epsilon_1} \cdot \vec{e_x}, \nonumber\\
   \tilde{J}_2 
   \approx \frac{e^2 a_1^2}{m^2 c^4} \left( \frac{36}{8} \frac{p_x^2}{m^2 c^2} \left( \vec{\epsilon_1} \cdot \vec{e_x} \right)^2 - \frac{3}{8}\vec{\epsilon}_1^{~2} \right).
  \end{eqnarray}
  Putting all this together, the $ \textit{S} $  matrix of Eq. 2 for small transverse momentum is estimated as follows
\begin{eqnarray}
& S\approx \dfrac{i}{2} T\dfrac{e^{3}a_1^{2}a_2}{m^{3}c^{6}}\left[ 
\dfrac{3}{8\hbar}p_x c \vec{\epsilon_1}^{2} \vec{\epsilon_2}^{*} .\vec{e_x}+
\dfrac{3}{4\hbar}p_x c \vec{\epsilon_1}.\vec{e_x} \vec{\epsilon_1}.\vec{\epsilon_2}^{*} \right.\nonumber\\
 &\left. -\dfrac{9i}{8}\omega \vec{\epsilon_1}^2 
 (\vec{\epsilon_2}^{*}\times\vec{e_z}). \vec{\sigma}\right]
  =\dfrac{i}{2}T\xi_1^2\xi_2\hat{\Omega}.
\end{eqnarray} 
Here $ \xi_{1,2}=\dfrac{ea_{1,2}}{mc^2}$ are the common dimensionless field amplitudes used in atomic physics. It is clear that for electrons entirely on $ z $ axis, the transition from $ p $ to $ p^\prime $ is dependent on the polarization beam vectors $  \vec{\epsilon_i} $ and the direction of electron spin along $z$ axis.
Generally for combination of the fundamental laser beam with higher harmonics laser beam, $ \hat{\Omega} $ has the common form presented in Eq. 7 and just the coefficients of each term are changed. For the electron with $ p_x=p_y=0 $, the third term of Eq. 7 has an essential role in the spin state of the scattered electron. As a result, we expect that the electron parallel to the beam axis in bichromatic ($\omega$:$3\omega$) laser beams shows the similar spin dynamics behavior as in bichromatic ($\omega$:$2\omega$) standing waves.

\section{Numerical Results}
In this section, the numerical results will be presented for the spin-dependent Kapitza-Dirac scattering in bichromatic counter-propagating laser fields with a frequency ratio of 3. Taking into account the combined vector potential and rewriting Dirac equation in momentum space, we find a system of coupled ordinary differential equations. The numerical solution of differential equations is obtained by employing a Crank-Nicholson scheme. The solutions are the absolute square values of the expansion coefficients that represent scattering probability of the electron in the particular quantum state. $ c_n^{\zeta}(n=0,\pm1,\pm2,...)$ coefficients represent electrons with momentum $p_n=(p_x,0,nk)$. The index $ \zeta\in\lbrace+\uparrow,+\downarrow,-\uparrow,-\downarrow \rbrace$ labels the sign of the energy and the spin direction. These states can be denoted by $\vert nk,\zeta\rangle $.

In the four-photon KD effect, the vector potential for the bichromatic field $ (\omega:3\omega) $ can be described in the form of 
\begin{equation}
\vec{A} =A_1\left[\cos (k(z\pm ct))\hat{\epsilon}_1\right]+A_2\left[\cos (3k(z\pm ct))\hat{\epsilon}_2\right],
\end{equation}
where $A_1$ and $A_2 $ are the amplitudes of standing waves and $\epsilon_1$ , $\epsilon_2$ are polarization vectors. The electron with initial longitudinal momentum of $p_z=-3k $ in presence of the mentioned vector potential is scattered into mirror mode with longitudinal momentum $p_z=+3k $ by exchanging $ 6k $  momentum. For all next calculations, we start with longitudinal momentum $p_z=-3k $ and spin projection either up or down, while the other momenta modes at initial time will be zero. As mentioned in Bragg regime, transfer of population is restricted to $p_z= +3k $ for the final state and occupation probability of other momenta modes is very small. When the electron has component perpendicular to laser beam direction, all four states $\vert-3,+\uparrow\rangle $, $\vert-3,+\downarrow\rangle$, $\vert+3,+\uparrow\rangle $ and $\vert+3,+\downarrow\rangle $ participate in the interaction. It is worth noting that because of focusing on the effect of $\vec{\sigma}.\vec{B}$ in this work, we choose electron momentum to be parallel to the axis of the laser beam, however practically the influence of $ \vec{p}.\vec{A} $ cannot be ignored.

In all simulations, smooth switching on and off laser fields are $sin^2 $ slopes for five cycle laser periods with a flat top function. To study spin effects in four-photon KD diffraction, the following various polarizations of this bichromatic standing wave are considered:
\begin{figure}[t]
\resizebox{86mm}{68mm}{\includegraphics{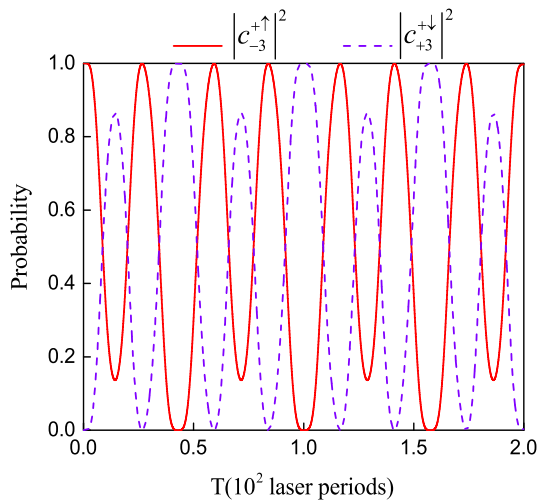}}
\caption{\label{fig:epsart}Time evolution of the occupation probability in four-photon KD effect with linear fundamental beam and counter propagating linear third harmonic. The overall laser intensity is $ 3 \times10^{22}$ W$\textnormal{cm}^{-2}$ with wavelength $\lambda=1.03$ nm, field parameters for beam $\omega$ and $3\omega$ are $A_1=18 \times10^{3}$ eV and $A_2=6 \times10^{3}$ eV respectively. The electron enters the laser fields with $p_z=-3k$ along the laser field direction.}
\end{figure} 
\begin{figure}[b]
\resizebox{86mm}{68mm}{\includegraphics{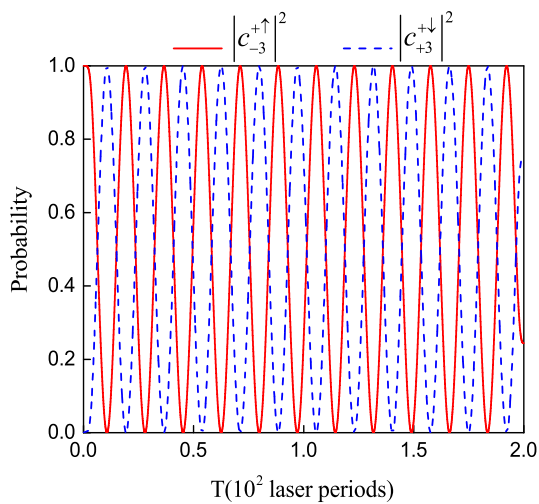}}
\caption{\label{fig:epsart} Rabi oscillation in both linear polarization setup for vector potential mentioned in Eq. 10. All other laser and electron parameters are same as Fig. 2.}
\end{figure}
\subsection{\label{sec:level2}Linear Polarization for Two Laser Beams (Lin-Lin)}

In the first setup, the fundamental field and its third harmonic are linearly polarized along the $ x $ axis
 \begin{equation}
\vec{A} =A_1\left[\cos(k(z-ct))\hat{e}_x\right]+A_2\left[\cos (3k(z+ct))\hat{e}_x\right].
\end{equation}
The incident electron is $-3\hbar k$ along the laser propagation direction and has no component parallel to the polarization direction. The overall laser intensity in Lin-Lin setup is $ I=\frac{\omega^2A_1^2+(3\omega)^{2}A_2^2}{8\pi c} $, when the amplitudes of beams are in maximum of their value. The angular momentum of fundamental laser beam $ \omega $ considered for all numerical solutions is $1.2\times10^3 $ eV. Fig. 2 presents the typical behavior of an electron in both linear laser beams. For the electron that is injected with spin up, a Rabi oscillation takes place between $\vert-3,+\uparrow\rangle $ and $\vert+3,+\downarrow\rangle $. The interaction in this field configuration is independent of the initial electron spin state such as the electron in initial momentum and the spin down state $\vert-3,+\downarrow\rangle $ is also scattered into the reflected momentum and spin up state $\vert+3,+\uparrow\rangle $ and the Rabi oscillation is similarly as Fig. 2. This symmetry of spin-flipping exists also in three-photon bichromatic$ (\omega:2\omega) $ KD effect with linear polarization [22].

 The shape of oscillation in bichromatic four-photon KD effect is sinusoidal and has two distinct peaks that are different in size in Fig. 2. It means that the probability of $ \vert c_{-3}^{+\uparrow}\vert^2 $ has sinusoidal oscillation whose minimum amplitude alternatively changes between 0.16 and 0.0 while $ \vert c_{+3}^{+\downarrow}\vert^2 $ oscillate similar to $ \vert c_{-3}^{+\uparrow}\vert^2 $ but it's maximum amplitude changes between 0.84 and 1.0. The Rabi oscillation period is about $ 2.8 $ fs and simulation shows that even with changing the standing waves amplitudes, these two distinct peaks for $ \vert c_{-3}^{+\uparrow}\vert^2 $ and $ \vert c_{+3}^{+\downarrow}\vert^2 $ will not be destroyed. 
 
The vector potential $ A(x)=f(t)[A_1(k_1.x)+A_2(k_2.x)] $ with slow envelope function $f(t)$ which was given in bichromatic $ (\omega:2\omega) $ KD effect demonstrates the typical Rabi oscillation with one peak [22]. Our results confirm that with similar vector potential for bichromatic $ (\omega:3\omega) $ KD effect
\begin{equation}
\vec{A} =f(t)\left( A_1\left[\cos(kz)\hat{e}_x\right]+A_2\left[\cos(3kz)\hat{e}_x\right] \right),
\end{equation} 
the typical oscillatory behavior with sinusoidal oscillation appears as shown in Fig. 3. The Rabi cycle is fully developed in Fig. 3 and the period of Rabi oscillation is $ 2\pi/\Omega_R=0.84 $ fs. By comparing vector potentials mentioned in Eq. 9 and Eq. 10, it's obvious that existence of $\cos(\omega_{i=1,2}t)$ part in numerical solutions results in different oscillation behavior.
  
\subsection{\label{sec:level2}Co-rotating Circular Fundamental and Third Harmonic Fields (Cir-Cir)}
We now look over an electron in two circular bichromatic counter propagating but co-rotating waves given by
\begin{eqnarray}
&\vec{A} =\frac{A_1}{\sqrt{2}}\left[\cos (k(z-ct))\left(\hat{ e}_x+i\hat{e}_y\right)\right]  \nonumber\\
 & +\frac{A_2}{\sqrt{2}}\left[\cos (3k(z+ct))\left(\hat{e}_x+i\hat{e}_y\right)\right]. 
\end{eqnarray}
If the electron is initially in spin up state with momentum $ -3k $, no diffraction occurs in this field configuration, as shown in upper panel of Fig. 4. In contrast for the electron with spin down and initial momentum $ -3k $ in lower panel, a Rabi oscillation between two states $\vert-3,+\downarrow\rangle $ and $\vert+3,+\uparrow\rangle $ takes place. This spin dependent diffraction behavior implies that it is possible to separate electrons based on their initial spin state in circularly polarized laser beams. The shape of Rabi oscillation in cir-cir setup for an electron with spin down is interesting. Still two peaks in oscillation exist and considering that maximum values of peaks are 1.0 and 0.96, the difference in maximum of peaks is less in this setup. As represented in Fig. 4 for two intervals about $0.56 $ fs the population of modes do not transfer and the electron maintains its momentum and spin. In fact similar to Fig. 2 the electron does not have instant changing momentum and spin. Also in this setup, the vector potential without $\cos(\omega_i t )$ part in Eq. 11 results in the typical sinusoidal Rabi oscillation similar to Fig. 3. 
  
\begin{figure}[t]
 \centering
\resizebox{86mm}{68mm}{\includegraphics{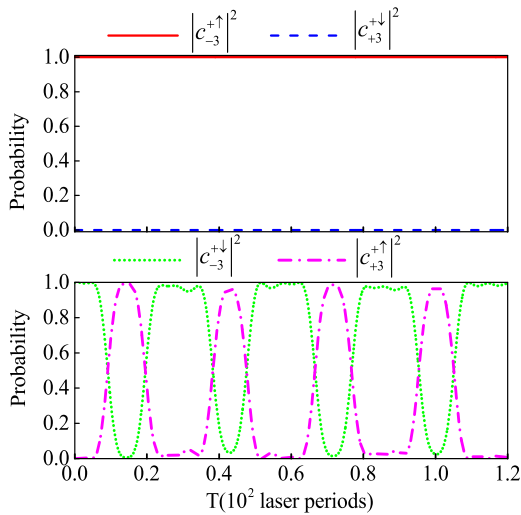}}
\caption{\label{fig:epsart}Time evolution of the occupation probability in four-photon KD effect with co-rotating circular bichromatic waves. The combined laser intensity is $ 7\times10^{22}$ W$\textnormal{cm}^{-2} $ with wavelength $\lambda=1.03$ nm, field amplitude for both beam is $A_1=A_2=30\times10^{3}$ eV. The electron enters the laser fields with $p_z=-3k$ along the laser direction. Upper panel of the figure shows that electron with spin up does not scatter in this setup.}
\end{figure}

\subsection{\label{sec:level2}Combination of Linear and Circular Polarization Fields (Lin-Cir)}
To examine the result of the last section about spin separation, we focus on a bichromatic setup in which the fundamental beam is linearly polarized and the third harmonic beam is circularly polarized
\begin{eqnarray}
&\vec{A} =A_1\left[\cos (k(z-ct))\left(\hat{ e}_x\right)\right]  \nonumber\\
 & +\frac{A_2}{\sqrt{2}}\left[\cos (3k(z+ct))\left(\hat{e}_x+i\hat{e}_y\right)\right].
\end{eqnarray}
By choosing the high frequency beam to be circularly polarized, as shown in Fig. 5, only the population probability starting from $\vert-3,+\uparrow\rangle $ travels to $\vert+3,+\downarrow\rangle$ state and a Rabi oscillation takes place. The Rabi period is $2.8$ fs and with the same parameters of the lin-lin setup, the population of modes does not transfer completely. The maximum of Rabi amplitude of $\vert+3,+\downarrow\rangle$ state only reaches 0.84.  When the electron spin is down and has $ -3k$ momentum $\vert-3,+\downarrow\rangle $, no scattering occurs at all. According to the $ \textit{S}$ matrix in Eq. 7 by putting $ \epsilon_1=\hat{ e}_x $ and $ \epsilon_2=(\hat{ e}_x+i\hat{ e}_y)/\sqrt{2} $ and due to existence $ \sigma_-$ in Rabi frequency $\hat{\Omega} $, the symmetry transition between the spin up $ \rightarrow $ down and spin down $ \rightarrow $ up disappears.
\begin{figure}[t]
 \centering
\resizebox{86mm}{68mm}{\includegraphics{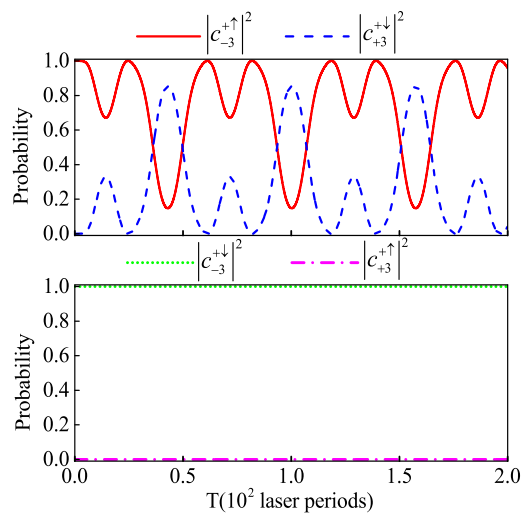}}
\caption{\label{fig:epsart}Time evolution of the occupation probability in four-photon KD effect with a hybrid setup that fundamental laser beam linear and the third harmonic is circularly polarized. The other simulation parameters are the same as in Fig. 2. In upper panel electron beam is initially spin up and spin filliping transition occur. The electron beam with spin down in lower panel is not diffracted.}
\end{figure}

 For the case with a fundamental frequency beam of circular polarization and a third harmonic beam of linear polarization, the spin flipping symmetry still exists, i.e. by choosing $ c_{-3}^{+\uparrow}(t=0)=1 $ or $ c_{-3}^{+\downarrow}(t=0)=1 $ the population is transferred to $\vert+3,+\downarrow\rangle $ and $\vert+3,+\uparrow\rangle $, respectively. We have a Rabi oscillation similar to upper panel of Fig. 5 for both electron spin states. In fact, the asymmetry of spin arises only for the circularly polarized high frequency laser beam. As the setup of both linearly polarized beams, if $ cos(\omega_i t)$ is not considered, we find a typical Rabi oscillation. 
\subsection{\label{sec:level2} Combination of Linear - Elliptical Polarization Fields (Lin-Ellip)}
When an elliptically polarized laser beam with frequency $ \omega $ and $ \epsilon_1=(2\hat{ e}_x+i\hat{ e}_y)/\sqrt{5}$ is combined with a linearly laser beam of frequency $ 3\omega $ and $\epsilon_2=\hat{ e}_x $, the Rabi oscillation for both up and down spin states of the electron occurs. Numerical results demonstrate well the spin-filliping symmetry. For electron with $ -3k$ momentum, whether the initial spin is up or down, we have a Rabi oscillation behavior as shown in Fig. 6 with dash line. For $ A_1=3A_2=18\times10^3 $ eV the  maximum of Rabi amplitude in this setup is $0.25 $.   

 In case of linear polarization for fundamental frequency and elliptical polarization for third harmonic, the vector potential is given by 
\begin{eqnarray}
&\vec{A} =A_1\left[\cos (k(z-ct))\left(\hat{ e}_x\right)\right]  \nonumber\\
 & +\frac{A_2}{\sqrt{5}}\left[\cos (3k(z+ct))\left(2\hat{e}_x+i\hat{e}_y\right)\right]. 
\end{eqnarray}
Solving Dirac equation for this vector potential gives a Rabi oscillation for the electron with initial spin up state as seen in Fig. 6 with solid line. Exactly the same exact Rabi oscillation will be happen if the spin of the initial electron is down. Therefore, with vector potential given by Eq. 13, the diffraction probability dose not depend on the spin orientation of the incident electrons, and with elliptical polarization for high harmonic, unlike circular polarization case, the electron can not be spin polarized. 
In fact the Rabi frequency obtained by $ S$ matrix in Eq. 7 also predicst the symmetry of spin-flipping behavior. The inequality of polarization amplitude brings out the existence of $ \sigma_x $ or $ \sigma_y $ in third term of Rabi frequency of Eq. 7. In this condition the separation of electrons due to their initial spin states vanishes.
\begin{figure}[t]
 \centering
\resizebox{86mm}{68mm}{\includegraphics{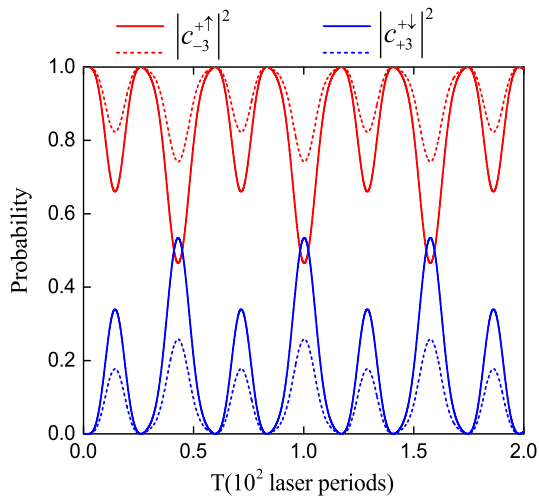}}
\caption{\label{fig:epsart}Time evolution of the occupation probability in four-photon KD effect from a linear polarized fundamental laser beam and third harmonic with $ (2\hat{ e}_x+i\hat{ e}_y)/\sqrt{5} $ polarized is shown by solid line. The dash line show the elliptically polarization of fundamental beam and linearly polarization of third harmonic beam. The fundamental laser wavelength $\lambda=1.03$nm and field amplitudes are the same as Fig. 2. For the incident electron with spin up or spin down, the Rabi transitions between two resonant states in any combination of Lin-Ellip configuration will be happen.}
\end{figure}

\section{Conclusion}		
Four-photon Kapitza-Dirac scattering occurs in the bichromatic standing wave when the electron beam absorbs three photons from the laser field with $ \omega $ frequency and emits one photon to the counter-propagation laser beam with $ 3\omega $ frequency. The initial electron spin and photon helicity are two factors that can affect on the polarization of the free electron in the bichromatic $(\omega:3\omega)$ KD effect. In this work, it is shown when the fundamental laser beam is linear and 3rd harmonic is circularly polarized, Rabi oscillation occurs only for electrons with spin up. When the initial electrons have spin down, diffraction does not exist and the symmetric flipping spin effect disappears.

In this study, we showed that the shape of Rabi frequency in bichromatic $(\omega:3\omega)$ vector potential has two unequal peaks that change with beam amplitude and the laser intensity. This different oscillation is due to fast time varying $\cos(\omega_i t )$ part in vector potential. Our results also indicate that these two peaks in the Rabi oscillation exist for other harmonics like second and forth harmonics. The Rabi frequency obtained from analytical $\textit{S}$ matrix is congruous with numerical results. Rabi frequency obtained from $\textit{S}$ matrix method predicts that even for other harmonics with circular polarization, the electron is diffracted based on its initial spin state. 

The momentum and spin of free electrons in four-photon bichromatic KD effect can be controlled by a suitable laser filed configuration. A spin-unpolarized incident electron beam with $ -3k$ momentum after interaction with a linear fundamental laser beam and third harmonic with circular polarization, splited into two portions $\vert-3,+\uparrow\rangle $ and $\vert+3,+\downarrow\rangle $ with Rabi oscillation. The electron with $+3k$ momentum is entirely spin polarized. By choosing the electron along the laser beams and circular polarization of the high frequency laser beam in bichromatic KD effect, the electron spin filter works for other harmonics participating in the Kapitza-Dirac effect.

\section{References}
\bibliography{p7}

\end{document}